\documentclass[11pt]{article}

\usepackage[letterpaper,margin=1in]{geometry}
\usepackage{amsmath}
\usepackage{bm}
\usepackage{amsfonts}
\usepackage{amssymb}
\usepackage{booktabs}
\usepackage{graphicx}
\usepackage{pgfplots}
\usepackage{subcaption}
\pgfplotsset{compat=1.18}
\usepackage[round]{natbib}
\usepackage[hidelinks]{hyperref}

\title{Weak Factor Recovery in Sparse Bayesian Factor Models under Exchangeable Shrinkage Process Prior}
\author{Margarita Grushanina\\
\small Department of Epidemiology and Biostatistics, School of Public Health,\\
\small Imperial College London}
\date{}

\begin{document}
\maketitle

\begin{abstract}
  Sparse Bayesian factor models provide a flexible framework for learning low-dimensional latent structure in multivariate data, but inference on the number of active factors remains challenging when some factors are weak or highly sparse. We study weak factor recovery under a variance based exchangeable shrinkage process (ESP) prior. We derive the posterior factor activation mechanism and show that, conditional on the global and local shrinkage parameters, activation is governed by the standardised loading energy together with the ESP prior activation odds. This yields an explicit activation boundary and clarifies how signal strength, sparsity, the number of observed variables, and the truncation level affect recovery. Simulation experiments show a clear transition between shrinkage and recovery around this boundary and demonstrate that sparsity can substantially affect activation even when raw loading energy is held fixed. They further show that factor activation and recovery of the relative loading pattern across variables can behave differently. These results clarify when weak latent factors become difficult to distinguish from redundant components under ESP shrinkage.
\end{abstract}

\noindent\textbf{Keywords:}
Sparse factor analysis; exchangeable shrinkage process prior; factor selection; weak factors

\section{Introduction}
\label{sec:Intro}

Latent factor models provide a parsimonious representation of the dependence structure in high-dimensional multivariate data by expressing variation in a large number of observed variables through a substantially smaller number of latent factors. In Bayesian factor analysis, sparsity inducing priors on the factor loadings can further regularise this representation by allowing individual factors to be associated with only subsets of the observed variables \citep{West2003, Carvalho2008, KnowlesGhahramani2011, FruehwirthSchnatterEtAl2025}. Such sparse factor structures are particularly useful when the latent components are expected to have an interpretable local structure, as is frequently the case in high-dimensional biomedical and genomic applications, where a latent factor may represent a biological mechanism involving only a relatively small subset of molecular features \citep{Carvalho2008, DeVito2021}.

A central challenge in factor analysis is determining the number of latent factors. Classical approaches include information criteria \citep{BaiNg2002} and eigenvalue based procedures, while Bayesian methods have relied on marginal likelihoods and posterior model probabilities to compare factor models of different dimensions \citep{Polasek1997}, as well as trans-dimensional sampling schemes such as reversible jump MCMC \citep{Green1995, LopesWest2004}. In particular, \citet{LopesWest2004} developed a tailored reversible-jump scheme in which preliminary fixed-dimension MCMC runs are used to construct more efficient proposals for between-model moves. Although these approaches provide a principled treatment of uncertainty in the factor dimension, repeated model fitting or trans-dimensional sampling can become computationally demanding, particularly in high-dimensional settings.

These computational challenges motivated the development of infinite factor models, in which the latent representation is allowed to be potentially unbounded and the effective number of factors is inferred through nonparametric priors. \citet{KnowlesGhahramani2011}, for example, used an Indian buffet process (IBP) prior to obtain a sparse nonparametric factor model in which the number of latent features is learned from the data. \citet{BhattacharyaDunson2011} instead introduced a multiplicative gamma process prior that imposes progressively stronger shrinkage on successive columns of the loading matrix, allowing an infinite factor representation to be adaptively truncated to a finite number of practically relevant factors. Related cumulative shrinkage process priors impose increasing shrinkage through a spike-and-slab construction in which the probability of assigning a factor to the spike increases with the factor index \citep{Legramanti2020}. A complementary approach is to use a finite but deliberately overfitting factor representation and shrink redundant factors towards zero. Exchangeable shrinkage process (ESP) priors provide such a construction while avoiding an explicit ordering of the factor-specific slab probabilities \citep{FruehwirthSchnatter2023}. This yields a computationally attractive framework in which an upper truncation $H$ is chosen larger than the expected factor dimension and the effective number of active factors is learned from the data.

The success of this strategy, however, depends not only on shrinking truly redundant factors, but also on retaining factors that are present in the data but provide comparatively weak evidence for activation. This distinction becomes particularly important in sparse high-dimensional settings. A factor may be weak because its non-zero loadings are small, because it is supported on only a small subset of variables, or through a combination of both. Moreover, in hierarchical sparse Bayesian factor models, factor activation is determined jointly by factor specific spike-and-slab shrinkage and by global and local shrinkage acting on the individual loadings. Consequently, factors with similar overall loading signal need not provide the same evidence for activation, and accurate recovery of the relative loading pattern across variables need not imply that the corresponding factor is classified as active. The effects of the number of observed variables $p$ and the truncation level $H$ are also not immediate, since both alter the balance between evidence supplied by the loading column and the prior preference for a parsimonious factor representation.

In this paper, we study weak factor recovery under a variance based ESP specification and make the posterior activation mechanism explicit. Conditional on the global and local shrinkage parameters, we show that the evidence supplied by a loading column is summarised by its standardised loading energy, while the ESP contributes the prior odds of factor activation. This yields an explicit activation boundary separating factors favoured by the spike from those favoured by the slab. The resulting representation allows us to characterise how signal strength, the number of observed variables $p$, and the truncation level $H$ affect factor activation, and to distinguish the direct role of standardised loading energy from the indirect effect of sparsity through the learned global and local shrinkage parameters. Simulation experiments show a clear transition between shrinkage and recovery around the activation boundary. They further show that factors with the same raw loading energy can have substantially different activation probabilities when their signal is distributed over supports of different sizes, and that recovery of the relative loading pattern across variables can remain accurate even when the corresponding factor is strongly favoured to belong to the spike. In the regimes considered, increasing $p$ improves recovery when support grows proportionally with dimension, whereas increasing $H$ raises the conditional activation threshold but does not necessarily reduce recovery because the hierarchical shrinkage parameters, and hence the standardised loading energy, adapt to the fitted model.

The rest of the paper is organised as follows. Section~\ref{sec:BayesFMandESP} introduces the sparse Bayesian factor model and the ESP prior. Section~\ref{sec:activationESP} derives the posterior factor activation mechanism and studies the roles of standardised loading energy, sparsity, the number of observed variables, and truncation in weak factor recovery. Section~\ref{sec:sim} presents the simulation study designed to investigate these theoretical effects empirically. Section~\ref{sec:concl} concludes with a brief discussion of the main findings and directions for future work.

\section{Sparse Bayesian factor model with exchangeable shrinkage prior}
\label{sec:BayesFMandESP}

\subsection{Sparse Bayesian factor model}
\label{sec:SBFM}

Let $\bm y_n=(y_{1n},\ldots,y_{pn})^\top$ denote the $p$-dimensional observation vector for observation $n=1,\ldots,N$. We consider the Gaussian factor model
\begin{align} \label{eq:BasicFM}
\bm y_n = \bm\Lambda \bm f_n + \bm\varepsilon_n,
\end{align}
where
\begin{align}
\bm f_n \sim N_{H^+}(\bm 0,\bm I_{H^+}), \qquad
\bm\varepsilon_n \sim N_p(\bm 0,\bm\Xi), 
\end{align}
with $\bm\Lambda=(\lambda_{ih})$ denoting a $p\times H^+$ loading matrix, $H^+$ the unknown number of latent factors, and
$\bm\Xi=\operatorname{diag}(\xi_1^2,\ldots,\xi_p^2)$
the diagonal matrix of idiosyncratic variances.

Sparse Bayesian factor models introduce regularisation through priors on the loading matrix $\bm\Lambda$, encouraging many of its elements to be small or exactly zero. Direct sparsity in the loading matrix can be induced through an IBP prior on the pattern of non-zero loadings, as in \citet{KnowlesGhahramani2011}. As another example, \citet{RockovaGeorge2016} combine the IBP representation with spike-and-slab LASSO shrinkage on the loading magnitudes. Alternatively, sparsity can be induced through hierarchical priors on latent variance parameters controlling the magnitude of individual loadings or entire loading columns \citep{BhattacharyaDunson2011, Legramanti2020}. A general scale mixture representation is
\begin{align*}
\lambda_{ih}\mid\psi_{ih}
\sim N(0,\psi_{ih}), \qquad
\psi_{ih} \sim P_\psi,
\end{align*}
where a shrinkage prior $P_\psi$ places substantial mass near zero thus inducing approximately sparse loading patterns. In either case, sparse loading patterns allow individual latent factors to be associated with only subsets of the observed variables.

\subsection{Exchangeable shrinkage process prior}
\label{sec:ESP}

When the number of factors $H^+$ is unknown, factor selection can be performed within an overfitting representation by fixing a truncation level $H>H^+$ and shrinking redundant loading columns towards zero. We therefore consider essentially the same representation as in Equation~\eqref{eq:BasicFM}
with the difference that now $\bm\Lambda=(\lambda_{ih})$ is a $p\times H$ loading matrix and
$$
\bm f_n\sim N_H(\bm 0,\bm I_H).
$$

For $i=1,\ldots,p$ and $h=1,\ldots,H$, we employ the three-level shrinkage hierarchy
\begin{align}   \label{eq:3sparsity}
\lambda_{ih}\mid\kappa,\theta_h,\phi_{ih}
\sim
N\!\left(0,\kappa\theta_h\phi_{ih}\right).
\end{align}
The global parameter $\kappa$ controls the overall magnitude of the loading matrix, the factor-specific parameter $\theta_h$ controls the magnitude of loading column $h$, and the local parameters $\phi_{ih}$ allow individual loadings within a column to undergo different degrees of shrinkage. While the priors for global and local shrinkage can take various forms, for conjugacy and computational convenience we assign
\begin{align*}
\kappa\sim IG(a_\kappa,b_\kappa),
\end{align*}
together with the local shrinkage hierarchy
\begin{align*}
\phi_{ih}\mid\beta_{\phi i}
\sim IG(a_\phi,\beta_{\phi i}), \qquad
\beta_{\phi i} \sim \mathrm{Gamma}(c_\phi,d_\phi).
\end{align*}

Automatic factor selection is induced by placing an exchangeable shrinkage process prior \citep{FruehwirthSchnatter2023} on the factor-specific variance parameters $\theta_h$. Thus, in the specification considered here, the ESP acts on a variance component shared by an entire loading column rather than directly on the individual loadings. We use the exchangeable continuous spike-and-slab construction
\begin{align}    \label{eq:ESPprior}
\theta_h\mid\tau_h
\sim
(1-\tau_h)IG(a_0,b_0)
+
\tau_h IG(a_1,b_1),
\end{align}
where $IG(a_0,b_0)$ denotes the spike component, concentrated on small values of $\theta_h$, and $IG(a_1,b_1)$ denotes the slab component. 

The inverse gamma specification is adopted to preserve conjugacy and
to obtain a closed form marginal distribution for the loading column
after integrating out $\theta_h$. Other ESP specifications are
possible, including alternative spike-and-slab distributions for the
factor-specific scale parameter or priors acting directly on the loadings. The theoretical and empirical results developed below should therefore be interpreted in the context of the variance based ESP specification adopted here. While the general competition between factor level evidence and exchangeable prior activation odds extends more broadly, the particular form of the standardised loading energy, the closed form conditional Bayes factor, and the resulting activation boundary derived in Section~\ref{sec:PosteriorActiv} depend on the Gaussian variance mixture and inverse gamma spike-and-slab construction.

The factor-specific slab probabilities are assigned exchangeable Beta priors,
\begin{align}    \label{eq:BetaTau}
\tau_h\mid\alpha,H
&\sim \mathrm{Beta}\!\left(\frac{\alpha}{H},1\right),
\qquad h=1,\ldots,H, \\ \nonumber
\alpha
&\sim \mathrm{Gamma}(a_\alpha,b_\alpha).
\end{align}

Introducing the latent allocation indicators $\gamma_h$, the ESP prior can equivalently be represented as
\begin{align}     \label{eq:BerGamma}
\gamma_h\mid\tau_h
&\sim \mathrm{Bernoulli}(\tau_h), \\ \nonumber
\theta_h\mid\gamma_h=g
&\sim IG(a_g,b_g),
\qquad g\in\{0,1\}.
\end{align}
Thus, $\gamma_h=0$ assigns $\theta_h$ to the spike component and identifies factor $h$ as inactive, whereas $\gamma_h=1$ assigns it to the slab and identifies factor $h$ as active.

Since the slab probabilities $\tau_h$ are exchangeable across the $H$ candidate factors (see \citealp{FruehwirthSchnatter2023}), the ESP prior does not impose an ordering on the loading columns. Marginalising over $\tau_h$ gives
\begin{align}    \label{eq:marg:prob:1:0}
\Pr(\gamma_h=1\mid\alpha,H)
&=
\frac{\alpha}{H+\alpha}, \\
\Pr(\gamma_h=0\mid\alpha,H)
&=
\frac{H}{H+\alpha},
\end{align}
and hence the prior odds of factor activation are
\begin{align}    \label{eq:priorOddsActive}
\frac{\Pr(\gamma_h=1\mid\alpha,H)}
{\Pr(\gamma_h=0\mid\alpha,H)}
=
\frac{\alpha}{H}.
\end{align}
The effective number of active factors is therefore represented by
$H^+=\sum_{h=1}^H\gamma_h$.

The resulting hierarchy separates three levels of shrinkage. At the factor level, the ESP prior on $\theta_h$, together with the allocation indicator $\gamma_h$, determines whether an entire loading column is assigned to the spike or the slab. The local parameters $\phi_{ih}$ control shrinkage of individual loadings within each column, while the global parameter $\kappa$ regulates the overall magnitude of the loading matrix. This separation is central to the weak factor recovery problem considered below, since the evidence for retaining a factor depends not only on its overall signal but also on how that signal interacts with the local and global shrinkage components.

\section{Activation of weak and sparse factors under the ESP prior}
\label{sec:activationESP}

\subsection{Posterior factor activation mechanism}
\label{sec:PosteriorActiv}

Posterior classification under ESP priors can generally be expressed by weighting the spike and slab marginal densities by their corresponding prior probabilities \citep{FruehwirthSchnatter2023}. Here we specialise this general allocation mechanism to the variance based ESP specification of Section~\ref{sec:ESP} and derive a representation that makes the evidence governing weak factor recovery explicit. Let
$\bm\lambda_h=(\lambda_{1h},\ldots,\lambda_{ph})^\top$
denote loading column $h$, and let
$\bm\phi_h=(\phi_{1h},\ldots,\phi_{ph})^\top$
denote the vector of element-wise shrinkage parameters corresponding to $1, \ldots, p$ elements of column $h$.

Conditional on the global and local shrinkage parameters and on the factor-specific variance $\theta_h$, the loading prior introduced in Section~\ref{sec:ESP} gives
$$
\bm\lambda_h\mid\theta_h,\kappa,\bm\phi_h
\sim
N_p\!\left(
\bm 0,
\theta_h\kappa\,\mathrm{diag}(\bm\phi_h)
\right).
$$
Equivalently, up to terms that do not depend on $\theta_h$,
$$
p(\bm\lambda_h\mid\theta_h,\kappa,\bm\phi_h)
\propto
\theta_h^{-p/2}
\exp\left\{
-\frac{1}{2\theta_h}
\sum_{i=1}^p
\frac{\lambda_{ih}^2}{\kappa\phi_{ih}}
\right\}.
$$
This motivates defining the standardised loading energy
\begin{align}
\label{eq:st:loading:energy}
S_h
=
\sum_{i=1}^p
\frac{\lambda_{ih}^2}{\kappa\phi_{ih}},
\end{align}
which summarises the loading column conditional on the global and local shrinkage parameters.

Taking the ESP prior as represented in Equation~\eqref{eq:BerGamma} and integrating out $\theta_h$ gives
$$
\bm\lambda_h
\mid
\gamma_h=g,\kappa,\bm\phi_h
\sim
t_{2a_g}\!\left(
\bm 0,
\frac{b_g}{a_g}
\kappa\,\mathrm{diag}(\bm\phi_h)
\right),
$$
where the second argument denotes the scale matrix. Thus, conditional on $\kappa$ and $\bm\phi_h$, integrating out the factor-specific variance yields spike-and-slab multivariate Student-$t$ distributions for the loading column, consistent with the marginal classification formulation of ESP priors in \citet{FruehwirthSchnatter2023}. Under the present variance based specification, both marginal densities depend on the loading column through the scalar quantity $S_h$. The corresponding marginal density under component $g$ is, up to terms common to both components,
$$
m_g(\bm\lambda_h)
\propto
\frac{
b_g^{a_g}\Gamma(a_g+p/2)
}{
\Gamma(a_g)
\left(b_g+S_h/2\right)^{a_g+p/2}
}.
$$

Consequently, the conditional Bayes factor comparing the slab to the spike, conditional on $\kappa$ and $\bm{\phi}_h$, can be written explicitly as a function of $S_h$ and $p$:
\begin{align}
\label{eq:Bayes:factor}
BF_h(S_h,p)
=
\frac{m_1(\bm\lambda_h)}
    {m_0(\bm\lambda_h)}
=
\frac{
b_1^{a_1}\Gamma(a_1+p/2)\Gamma(a_0)
}{
b_0^{a_0}\Gamma(a_0+p/2)\Gamma(a_1)
}
\times
\frac{
\left(b_0+S_h/2\right)^{a_0+p/2}
}{
\left(b_1+S_h/2\right)^{a_1+p/2}
}.
\end{align}

Combining the prior activation odds in Equation~\eqref{eq:priorOddsActive} with the conditional Bayes factor gives the posterior factor activation odds. Let $O_h$ denote the posterior odds that factor $h$ belongs to the slab rather than the spike:
\begin{align}
\label{eq:post:act:odds}
O_h
&=
\frac{
\Pr(\gamma_h=1\mid
\bm\lambda_h,\kappa,\bm\phi_h,\alpha,H)
}{
\Pr(\gamma_h=0\mid
\bm\lambda_h,\kappa,\bm\phi_h,\alpha,H)
}
=
\frac{\alpha}{H}BF_h(S_h,p).
\end{align}
Equivalently, on the log scale,
\begin{align}
\label{eq:post:act:logodds}
\log O_h
=
\log BF_h(S_h,p)
-
\log\left(\frac{H}{\alpha}\right).
\end{align}

Thus, conditional on the global and local shrinkage parameters, the loading column enters the posterior activation mechanism only through the scalar standardised loading energy $S_h$. While the general spike-and-slab allocation mechanism follows from the ESP construction, this reduction makes it possible to characterise explicitly the amount of standardised loading energy required for activation. The resulting decomposition provides the basis for deriving the activation boundary and for studying how signal strength, the number of observed variables $p$, the truncation level $H$, and the sparsity of the loading pattern affect the recovery of weak factors.

\subsection{Activation boundary and dimensional effects}
\label{sec:ActivationBoundary}

The posterior odds derived in Section~\ref{sec:PosteriorActiv} provide a natural criterion for factor activation. Factor $h$ is more likely to belong to the slab than to the spike whenever
$O_h>1$, or, as follows from Equation~\eqref{eq:post:act:logodds},
$$
\log BF_h(S_h,p)
>
\log\left(\frac{H}{\alpha}\right).
$$
Thus, the boundary between spike and slab allocation is determined by the balance between the evidence provided by the loading column and the prior penalty for activating an additional factor. When $BF_h(S_h,p)$ is strictly increasing in $S_h$ and $BF_h(0,p) < H/\alpha$, the corresponding activation threshold $T$ can be defined implicitly by
$$
\log BF_h(T,p)
=
\log\left(\frac{H}{\alpha}\right),
$$
so that factors with $S_h>T$ are favoured by the slab, whereas factors with $S_h<T$ are favoured by the spike.

The dependence of the Bayes factor on the standardised loading energy follows directly from Equation~\eqref{eq:Bayes:factor}. Differentiating with respect to $S_h$ gives
$$
\frac{\partial}{\partial S_h}\log BF_h(S_h,p)
=
\frac{1}{2}
\left\{
\frac{a_0+p/2}{b_0+S_h/2}
-
\frac{a_1+p/2}{b_1+S_h/2}
\right\}.
$$
Hence, for $a_0>a_1$, the Bayes factor is increasing in $S_h$ whenever
$$
S_h
>
\frac{
2\left[
(a_1+p/2)b_0-(a_0+p/2)b_1
\right]
}{
a_0-a_1
}.
$$
In particular, if
$$
(a_0+p/2)b_1
\geq
(a_1+p/2)b_0,
$$
the Bayes factor is increasing for all $S_h\geq0$. When this condition is not satisfied, the Bayes factor is initially
decreasing in $S_h$ and becomes increasing only above the critical
value
$$
S^* =
\frac{
2\left[(a_1+p/2)b_0-(a_0+p/2)b_1\right]
}{
a_0-a_1
}.
$$
This turning point should be distinguished from the activation
threshold $T$, which is determined by the posterior odds crossing
one. In particular, when $BF_h(0,p)<H/\alpha$, the activation
threshold exists and is unique on the increasing branch of the
Bayes factor, with $S_h>T$ favouring slab allocation and
$S_h<T$ favouring spike allocation.

The truncation level $H$ has a particularly transparent effect on factor activation. From Equation~\eqref{eq:post:act:logodds},
increasing $H$ reduces the posterior log odds of activation through an additive prior penalty. For two truncation levels $H_1$ and $H_2$, conditional on the remaining quantities,
$$
\log O_h(H_2)-\log O_h(H_1)
=
-\log\left(\frac{H_2}{H_1}\right).
$$
Thus, enlarging the overfitting representation raises the amount of evidence required for a candidate factor to be classified as active, with the direct effect of $H$ entering logarithmically through the ESP prior odds.

The role of the number of variables $p$ is different. Unlike $H$, $p$ enters the Bayes factor itself through both the normalising constants and the exponents of the spike and slab densities. Writing the log Bayes factor explicitly,
\begin{align*}
\log BF_h(S_h,p)
&=
a_1\log b_1-a_0\log b_0
+\log\Gamma(a_1+p/2)
-\log\Gamma(a_0+p/2)
+\log\Gamma(a_0)
\\
&-\log\Gamma(a_1)
+\left(a_0+\frac{p}{2}\right)
\log\left(b_0+\frac{S_h}{2}\right)
-\left(a_1+\frac{p}{2}\right)
\log\left(b_1+\frac{S_h}{2}\right).
\end{align*}
Consequently, there is no universal dimensional penalty associated with increasing $p$. Rather, for fixed $S_h$, the effect of $p$ depends jointly on the spike-and-slab parameters and on the standardised loading energy. Treating $p$ as continuous for interpretative purposes,
\begin{align*}
\frac{\partial}{\partial p}\log BF_h(S_h,p)
=
\frac{1}{2}\Bigg[
\psi\left(a_1+\frac{p}{2}\right)
-
\psi\left(a_0+\frac{p}{2}\right)
+
\log\left(
\frac{b_0+S_h/2}{b_1+S_h/2}
\right)
\Bigg],
\end{align*}
where $\psi(\cdot)$ denotes the digamma function. Since the sign of this derivative is not fixed, increasing $p$ may either strengthen or weaken the evidence for factor activation. In practice, however, the standardised loading energy $S_h$ may itself change with $p$, depending on how the factor support and loading magnitudes scale with dimension. This motivates examining the joint effects of dimension, signal strength, and sparsity.

\subsection{Signal strength, sparsity, and hierarchical shrinkage}
\label{sec:Sparsity}

The standardised loading energy $S_h$ combines two distinct
characteristics of a latent factor: the number of variables
on which it is active and the magnitude of its loadings.
To make this distinction explicit, consider an idealised
sparse loading column with support $\mathcal A_h$ and let
$m_h=|\mathcal A_h|$
denote the number of non-zero loadings. Define the average
standardised loading strength over the active support as
$$
\bar s_h
=
\frac{1}{m_h}
\sum_{i\in\mathcal A_h}
\frac{\lambda_{ih}^2}{\kappa\phi_{ih}}.
$$
Then
\begin{equation}
\label{eq:signal:supp}
S_h=m_h\bar s_h.
\end{equation}
Under the continuous shrinkage prior, individual loadings
are not exactly zero, so this decomposition applies
approximately when $\mathcal A_h$ is interpreted as the
effective support comprising variables with substantial
contributions to the standardised loading energy.
Hence, a factor may provide weak evidence for activation
because it is supported on only a small number of
variables, because its active loadings have small average
strength, or through a combination of both.
Sparsity and weakness therefore describe distinct
features of the latent structure.

The decomposition in Equation~\eqref{eq:signal:supp} helps
distinguish the effects of increasing the number of observed
variables from those of changing the sparsity of a factor.
Consider first how the standardised loading energy scales
with $p$. If the support proportion remains fixed, so that
$m_h=\rho p$, and the average standardised loading strength
$\bar s_h$ remains approximately constant, then $S_h$
grows proportionally with $p$. In contrast, if the support
size $m_h$ remains fixed, then $S_h$ remains of the same
order provided that $\bar s_h$ is stable. Thus, increasing
the number of variables need not increase the aggregate
standardised loading energy; this depends on how the
factor support scales with dimension.

A different question concerns the effect of sparsity at a
fixed dimension $p$. If the total standardised loading
energy $S_h$ is held fixed while the support size $m_h$
varies, then $\bar s_h=S_h/m_h$. Hence, a sparser factor
must have greater average standardised loading strength
over its active variables. However, as follows from
Equation~\eqref{eq:Bayes:factor}, such differences in the
distribution of loading energy do not affect the Bayes
factor directly when $S_h$ and $p$ are fixed.

An important consequence of the posterior activation mechanism
is that, conditional on $\kappa$ and $\boldsymbol{\phi}_h$,
the Bayes factor $BF_h(S_h,p)$ depends on the loading column
only through $S_h$, rather than on $m_h$ and $\bar s_h$
separately. Thus, for fixed $p$, two loading columns with the
same standardised loading energy provide identical direct
spike versus slab evidence, regardless of how that energy is
distributed across variables. In the full hierarchical model,
however, $\kappa$ and the local scales $\phi_{ih}$ are themselves
learned from the data. Concentrating the signal on a small
number of variables can therefore produce a different pattern
of local shrinkage than distributing it more diffusely, while
the global parameter $\kappa$ adjusts to the overall magnitude
of the loading matrix. Consequently, even when the raw loading
energy is held fixed, changes in sparsity may alter the learned
shrinkage parameters and hence the standardised energy $S_h$.
Sparsity therefore has no additional direct effect on factor
activation conditional on $S_h$, but can influence activation
indirectly through the hierarchical shrinkage mechanism.
This motivates the simulation experiment in which the support
size is varied while the raw loading energy is held fixed.

\section{Simulation study}
\label{sec:sim}

The simulation study investigates the recovery of weak latent factors under changes in signal strength, sparsity, the number of observed variables, and the truncation level. The experiments are constructed around a common baseline configuration and are designed to assess how the empirical recovery behaviour relates to the theoretical mechanisms described in the preceding sections.

\subsection{Simulation design}
\label{sec:SimulationDesign}

We consider a baseline Gaussian factor model as in
Equation~\eqref{eq:BasicFM} with $H_0^+=5$ true latent factors,
$p=100$ observed variables, and $N=200$ observations,
\begin{align*}
\bm y_n
&=
\bm\Lambda_0 \bm f_n + \bm\varepsilon_n,
\qquad n=1,\ldots,N,\\
\bm f_n
&\sim
N_5(\bm 0,\bm I_5),
\qquad
\bm\varepsilon_n
\sim
N_p(\bm 0,\bm\Xi_0),
\end{align*}
where $\bm\Xi_0=\bm I_p$. Each column of $\bm\Lambda_0$ has baseline
support fraction $0.20$, corresponding to $m_h=20$ active variables.
A common block comprising $40\%$ of the active support is shared by all five factors, while the remaining active positions are factor specific. On the active support, unscaled loading magnitudes are drawn independently from $U(0.8,1.2)$ with independent random signs, and the resulting loading columns are scaled by factor-specific signal strengths
$$
(c_1,\ldots,c_5)
=
(1.50,\,1.25,\,1.00,\,0.50,\,0.35).
$$
Factors $1$--$4$ are chosen to be clearly identifiable and serve as
stable reference factors throughout the simulation study. Factor $5$
is deliberately chosen to lie near the weak factor recovery boundary
and is the target of the recovery analysis below. Unless otherwise
stated, the model is fitted with truncation level $H=10$ using the ESP hierarchy of Section~\ref{sec:ESP}. All prior hyperparameters are held fixed across simulation configurations. The element-wise shrinkage hierarchy uses
$$
a_\phi=1.5,
\qquad
c_\phi=1.5,
\qquad
d_\phi=1.
$$

Posterior loading columns are matched to the true factors draw by draw using a one-to-one assignment that maximises the total absolute loading correlation across the true factors. For true factor $5$, let $h_5$ denote its matched posterior column at a given draw and define the loading recovery measure
$$
R_5
=
\left|
\operatorname{cor}
\left(
\bm\lambda_{05},
\bm\lambda_{h_5}
\right)
\right|,
$$
which provides a sign- and permutation-invariant measure of loading recovery. The activation indicator $\gamma_5$ and standardised loading energy $S_5$ are taken from the same matched posterior column. Within each simulated dataset, we report the posterior activation probability
$$
\pi_5=\Pr(\gamma_5=1\mid Y),
$$
where $Y$ denotes the observed data, together with the posterior medians of $R_5$ and $S_5$. We additionally report
$$
\Pr(H^+=H_0^+\mid Y),
\qquad
H^+=\sum_{h=1}^H\gamma_h,
$$
where $H_0^+$ denotes the true effective factor dimension. For the null configuration $c_5=0$, the fifth factor is absent and $H_0^+=4$. In this case we report $\Pr(H^+=4\mid Y)$ rather than matched factor $5$ quantities. For each retained posterior draw, the activation threshold $T$ is obtained by solving
$$
\log BF_h(T,p)
=
\log\left(\frac{H}{\alpha}\right)
$$
using the draw-specific value of $\alpha$. Within each simulated dataset, $T$ is summarised by its posterior median.

Starting from the baseline configuration, we vary one design parameter at a time. Factors $1$--$4$ retain their signal strengths throughout, only the strength or sparsity of factor $5$ is directly perturbed. As the support fraction of factor $5$ is varied, approximately $40\%$ of its active variables are selected from variables used by factors $1$--$4$ and the remainder from variables outside their supports. The resulting loading vector is rescaled to preserve its baseline raw loading energy exactly, with $\rho_5=0.20$ reproducing the baseline configuration. In the dimension experiment, the same generating mechanism, signal strengths, support proportions, and overlap proportion are retained as $p$ changes. In the truncation experiment, we consider values of $H$ that provide a genuinely overfitting representation relative to the true dimension $H_0^+=5$, and therefore exclude truncation levels only marginally larger than the true number of factors. Table~\ref{tab:simulation_design} summarises the four simulation regimes and the theoretical mechanism targeted by each.

\begin{table*}[t]
\centering
\footnotesize
\setlength{\tabcolsep}{7pt}
\caption{Simulation regimes varying one parameter at a time. Baseline values are shown in bold.}
\label{tab:simulation_design}
\begin{tabular}{p{2.9cm}p{5.2cm}p{7.2cm}}
\toprule
Regime & Grid (baseline in bold) & Tests \\
\midrule

Signal strength $c_5$
&
$0,\,0.20,\,0.25,\,0.30,\,\mathbf{0.35},\,0.40,\,0.50$
&
Transition from shrinkage to recovery as the fifth factor signal
increases, and its relation to the activation boundary in $S_5$
(Sec.~\ref{sec:ActivationBoundary})
\\[5pt]

Sparsity of factor $5$, $\rho_5=m_5/p$
&
$0.05,\,0.10,\,\mathbf{0.20},\,0.30,\,0.40$;
non-zero loadings are rescaled so that
$\|\bm\lambda_{05}\|^2$ equals its baseline value
&
Whether redistributing the same raw loading signal over different numbers of variables changes posterior $S_5$ and activation through the local and global shrinkage hierarchy
(Sec.~\ref{sec:Sparsity})
\\[5pt]

Number of observed variables $p$
&
$50,\,\mathbf{100},\,200,\,400$;
support fraction fixed at $0.20$
&
Combined effect of $p$ on the conditional Bayes factor and activation threshold,
and on the amount of available signal when support size grows
proportionally with $p$
(Secs.~\ref{sec:ActivationBoundary} and~\ref{sec:Sparsity})
\\[5pt]

Truncation level $H$
&
$8,\,\mathbf{10},\,15,\,20,\,30$
&
Logarithmic effect of $H$ on posterior activation odds and the
corresponding shift in the activation threshold $T$
(Sec.~\ref{sec:ActivationBoundary})
\\

\bottomrule
\end{tabular}
\end{table*}

We use $R=20$ independent simulation replications. Within each replication, common simulated quantities are reused across configurations whenever they are not themselves varied by the experiment, so that comparisons within a regime isolate the parameter being varied.
The baseline configuration is shared across the four regimes. The prior hyperparameters used for all simulations are reported in the
Supplementary Material, Section~\ref{app:SimulationHyperparameters}.

Posterior inference was performed using four independently seeded MCMC chains. Each chain was run for $10{,}000$ iterations, with the first $5{,}000$ iterations discarded as burn-in and every fifth subsequent draw retained, yielding $1{,}000$ retained draws per chain and $4{,}000$ posterior draws per fitted model. Sampling behaviour was assessed using the rank-normalised split-$\widehat R$, bulk effective sample size (bulk-ESS), and tail effective sample size (tail-ESS) diagnostics of \citet{vehtari2021rank}, as implemented in the R package \texttt{posterior} \citep{buerkner2020posterior}. We monitored the global shrinkage parameter $\kappa$, the ESP concentration parameter $\alpha$, and the effective factor dimension $H^+$. When the fifth
factor was present, its matched activation indicator $\gamma_5$ and
standardised loading energy $S_5$ were additionally monitored.
Following \citet{vehtari2021rank}, $\widehat R<1.01$ and
bulk-ESS of at least $400$ were used as diagnostic benchmarks, with
tail-ESS of at least $400$ additionally considered for continuous
quantities. Detailed definitions, computation, and diagnostic results are reported in the Supplementary Material,
Section~\ref{app:MCMCDiagnostics}.

\subsection{Simulation results}
\label{sec:SimulationResults}

Table~\ref{tab:simulation_results} summarises weak factor activation,
loading recovery, standardised loading energy, and recovery of the
effective factor dimension across the four simulation experiments.
Entries are medians and interquartile ranges over $20$ independently
simulated datasets.

\begin{table*}[t]
\centering
\footnotesize
\setlength{\tabcolsep}{3pt}
\caption{Weak factor recovery across the simulation experiments.
Entries are medians [interquartile ranges] across $20$ datasets.
Baseline values are shown in bold. The common baseline configuration is
$c_5=0.35$, $\rho_5=0.20$, $p=100$, and $H=10$; each experiment varies
one of these quantities while holding the others fixed. For $c_5=0$,
factor $5$ is absent and the true dimension is $H_0^+=4$; otherwise
$H_0^+=5$.}
\label{tab:simulation_results}
\begin{tabular}{llccccc}
\toprule
Regime & Setting & $\pi_5$ & $R_5$ & $S_5$ & $T$ &
$\Pr(H^+=H_0^+\mid\bm Y)$ \\
\midrule

Signal
& $c_5=0$
& -- & -- & --
& $10.8\,[10.8,\,10.8]$
& $0.988\,[0.979,\,0.991]$ \\

& $c_5=0.20$
& $0.003\,[0.001,\,0.003]$
& $0.335\,[0.303,\,0.366]$
& $4.53\,[4.47,\,4.60]$
& $10.8\,[10.8,\,10.8]$
& $0.007\,[0.004,\,0.010]$ \\

& $c_5=0.25$
& $0.014\,[0.011,\,0.027]$
& $0.489\,[0.441,\,0.522]$
& $5.12\,[5.01,\,5.32]$
& $10.8\,[10.8,\,10.8]$
& $0.018\,[0.015,\,0.031]$ \\

& $c_5=0.30$
& $0.256\,[0.195,\,0.372]$
& $0.691\,[0.641,\,0.728]$
& $6.95\,[6.64,\,7.94]$
& $10.8\,[10.8,\,10.8]$
& $0.257\,[0.189,\,0.365]$ \\

& $c_5=\mathbf{0.35}$
& $0.860\,[0.805,\,0.914]$
& $0.807\,[0.776,\,0.831]$
& $18.4\,[17.6,\,19.6]$
& $10.7\,[10.7,\,10.7]$
& $0.849\,[0.775,\,0.893]$ \\

& $c_5=0.40$
& $0.973\,[0.955,\,0.988]$
& $0.860\,[0.839,\,0.879]$
& $22.0\,[21.1,\,22.5]$
& $10.7\,[10.7,\,10.7]$
& $0.958\,[0.933,\,0.970]$ \\

& $c_5=0.50$
& $0.998\,[0.996,\,0.999]$
& $0.917\,[0.904,\,0.925]$
& $26.7\,[25.8,\,27.3]$
& $10.7\,[10.7,\,10.7]$
& $0.981\,[0.970,\,0.988]$ \\[2pt]

\midrule

Sparsity
& $\rho_5=0.05$
& $0.001\,[0.001,\,0.003]$
& $0.922\,[0.911,\,0.930]$
& $4.61\,[4.50,\,4.76]$
& $10.8\,[10.8,\,10.8]$
& $0.004\,[0.004,\,0.006]$ \\

& $\rho_5=0.10$
& $0.072\,[0.049,\,0.098]$
& $0.882\,[0.861,\,0.890]$
& $6.10\,[5.94,\,6.33]$
& $10.8\,[10.8,\,10.8]$
& $0.074\,[0.052,\,0.100]$ \\

& $\rho_5=\mathbf{0.20}$
& $0.860\,[0.805,\,0.914]$
& $0.807\,[0.776,\,0.831]$
& $18.4\,[17.6,\,19.6]$
& $10.7\,[10.7,\,10.7]$
& $0.849\,[0.775,\,0.893]$ \\

& $\rho_5=0.30$
& $0.982\,[0.956,\,0.992]$
& $0.776\,[0.749,\,0.797]$
& $23.1\,[21.7,\,24.5]$
& $10.7\,[10.7,\,10.7]$
& $0.957\,[0.929,\,0.978]$ \\

& $\rho_5=0.40$
& $0.998\,[0.977,\,0.999]$
& $0.758\,[0.705,\,0.776]$
& $27.0\,[23.5,\,29.1]$
& $10.7\,[10.7,\,10.7]$
& $0.965\,[0.944,\,0.982]$ \\[2pt]

\midrule

Dimension
& $p=50$
& $0.082\,[0.025,\,0.142]$
& $0.577\,[0.517,\,0.639]$
& $2.74\,[2.52,\,2.94]$
& $5.53\,[5.53,\,5.54]$
& $0.083\,[0.038,\,0.122]$ \\

& $p=\mathbf{100}$
& $0.860\,[0.805,\,0.914]$
& $0.807\,[0.776,\,0.831]$
& $18.4\,[17.6,\,19.6]$
& $10.7\,[10.7,\,10.7]$
& $0.849\,[0.775,\,0.893]$ \\

& $p=200$
& $0.996\,[0.989,\,0.999]$
& $0.833\,[0.826,\,0.849]$
& $43.4\,[40.6,\,47.0]$
& $21.2\,[21.2,\,21.2]$
& $0.996\,[0.989,\,0.998]$ \\

& $p=400$
& $1.000\,[1.000,\,1.000]$
& $0.852\,[0.847,\,0.860]$
& $104\,[98.9,\,108]$
& $42.0\,[42.0,\,42.0]$
& $1.000\,[1.000,\,1.000]$ \\[2pt]

\midrule

Truncation
& $H=8$
& $0.617\,[0.500,\,0.748]$
& $0.809\,[0.776,\,0.833]$
& $13.8\,[10.6,\,16.3]$
& $10.6\,[10.6,\,10.6]$
& $0.601\,[0.487,\,0.728]$ \\

& $H=\mathbf{10}$
& $0.860\,[0.805,\,0.914]$
& $0.807\,[0.776,\,0.831]$
& $18.4\,[17.6,\,19.6]$
& $10.7\,[10.7,\,10.7]$
& $0.849\,[0.775,\,0.893]$ \\

& $H=15$
& $0.976\,[0.952,\,0.986]$
& $0.804\,[0.776,\,0.830]$
& $22.8\,[22.2,\,23.7]$
& $11.0\,[11.0,\,11.1]$
& $0.959\,[0.929,\,0.971]$ \\

& $H=20$
& $0.994\,[0.984,\,0.997]$
& $0.803\,[0.775,\,0.829]$
& $26.0\,[25.1,\,27.1]$
& $11.3\,[11.3,\,11.3]$
& $0.971\,[0.963,\,0.974]$ \\

& $H=30$
& $0.999\,[0.998,\,1.000]$
& $0.801\,[0.774,\,0.827]$
& $30.7\,[29.6,\,31.8]$
& $11.6\,[11.6,\,11.6]$
& $0.958\,[0.952,\,0.964]$ \\

\bottomrule
\end{tabular}
\end{table*}

\paragraph{Signal strength.}
The first experiment shows a clear transition from shrinkage to
recovery as the strength of factor $5$ increases. For
$c_5\leq0.25$, the factor is almost always assigned to the spike,
with median activation probabilities of $0.003$ at $c_5=0.20$ and
$0.014$ at $c_5=0.25$. At $c_5=0.30$, activation becomes
appreciable but remains uncertain, with $\pi_5=0.256$ and
$\Pr(H^+=5\mid\bm Y)=0.257$. Activation increases substantially at
$c_5=0.35$, where the median reaches $0.860$, and becomes nearly
certain for $c_5\geq0.40$. Loading recovery follows the same overall
pattern, with $R_5$ increasing from $0.335$ at $c_5=0.20$ to $0.917$
at $c_5=0.50$.

\begin{figure*}[t]
\centering

\caption{
Main weak factor recovery mechanisms.
Panel (a) shows the transition in posterior activation as the
standardised loading energy crosses the activation threshold
$T\approx10.8$ in the signal strength experiment; the horizontal
dotted line marks $\pi_5=0.5$.
Panel (b) shows that, although raw loading energy is held fixed,
the standardised loading energy increases with the support fraction
and crosses the activation threshold.
Panel (c) shows that, over the same sparsity experiment, factor
activation increases while loading recovery decreases.
Points show medians across $20$ simulated datasets.
}
\label{fig:weak_factor_recovery}

\vspace{0.4em}
\captionsetup[subfigure]{font=small,skip=2pt}

\begin{subfigure}[t]{0.30\textwidth}
\centering
\begin{tikzpicture}
\begin{axis}[
    width=\textwidth,
    height=0.78\textwidth,
    xlabel={$S_5$},
    ylabel={$\pi_5$},
    xmin=3,
    xmax=29,
    ymin=0,
    ymax=1.05,
    xtick={5,10,15,20,25},
    ytick={0,0.2,0.4,0.6,0.8,1.0},
    tick align=outside,
    tick label style={font=\scriptsize},
    label style={font=\small},
    axis x line*=bottom,
    axis y line*=left,
    ymajorgrids=true,
    grid style={dotted},
]

\addplot+[
    only marks,
    mark=*,
    mark size=2.3pt
]
coordinates {
    (4.53,0.003)
    (5.12,0.014)
    (6.95,0.256)
    (18.4,0.860)
    (22.0,0.973)
    (26.7,0.998)
};

\addplot[
    dashed,
    thick
]
coordinates {
    (10.8,0)
    (10.8,1.05)
};

\addplot[
    dotted,
    thick
]
coordinates {
    (3,0.5)
    (29,0.5)
};

\node[
    font=\scriptsize,
    anchor=north west
] at (axis cs:11.2,1.01) {$T$};

\end{axis}
\end{tikzpicture}

\caption{Activation boundary.}
\label{fig:activation_boundary}
\end{subfigure}
\hfill
\begin{subfigure}[t]{0.30\textwidth}
\centering
\begin{tikzpicture}
\begin{axis}[
    width=\textwidth,
    height=0.78\textwidth,
    xlabel={$\rho_5$},
    ylabel={$S_5$},
    xmin=0.025,
    xmax=0.425,
    ymin=0,
    ymax=30,
    xtick={0.05,0.10,0.20,0.30,0.40},
    xticklabels={$0.05$,$0.10$,$0.20$,$0.30$,$0.40$},
    ytick={0,5,10,15,20,25,30},
    tick align=outside,
    tick label style={font=\scriptsize},
    label style={font=\small},
    axis x line*=bottom,
    axis y line*=left,
    ymajorgrids=true,
    grid style={dotted},
]

\addplot+[
    mark=*,
    mark size=2.3pt,
    thick
]
coordinates {
    (0.05,4.61)
    (0.10,6.10)
    (0.20,18.4)
    (0.30,23.1)
    (0.40,27.0)
};

\addplot[
    dashed,
    thick
]
coordinates {
    (0.025,10.8)
    (0.425,10.8)
};

\node[
    font=\scriptsize,
    anchor=south east
] at (axis cs:0.415,10.9) {$T$};

\end{axis}
\end{tikzpicture}

\caption{Sparsity at fixed raw energy.}
\label{fig:sparsity_energy}
\end{subfigure}
\hfill
\begin{subfigure}[t]{0.33\textwidth}
\centering
\begin{tikzpicture}
\begin{axis}[
    width=\textwidth,
    height=0.78\textwidth,
    xlabel={$\rho_5$},
    ylabel={},
    xmin=0.025,
    xmax=0.425,
    ymin=0,
    ymax=1.05,
    xtick={0.05,0.10,0.20,0.30,0.40},
    xticklabels={$0.05$,$0.10$,$0.20$,$0.30$,$0.40$},
    ytick={0,0.2,0.4,0.6,0.8,1.0},
    tick align=outside,
    tick label style={font=\scriptsize},
    label style={font=\small},
    axis x line*=bottom,
    axis y line*=left,
    ymajorgrids=true,
    grid style={dotted},
    legend style={
        draw=none,
        fill=none,
        font=\scriptsize,
        at={(0.97,0.08)},
        anchor=south east,
        legend columns=2,
        /tikz/every even column/.append style={column sep=0.6em}
    },
]

\addplot+[
    mark=*,
    mark size=2.3pt,
    thick
]
coordinates {
    (0.05,0.001)
    (0.10,0.072)
    (0.20,0.860)
    (0.30,0.982)
    (0.40,0.998)
};
\addlegendentry{$\pi_5$}

\addplot+[
    mark=triangle*,
    mark size=3.1pt,
    dashed,
    thick
]
coordinates {
    (0.05,0.922)
    (0.10,0.882)
    (0.20,0.807)
    (0.30,0.776)
    (0.40,0.758)
};
\addlegendentry{$R_5$}

\end{axis}
\end{tikzpicture}

\caption{Activation and loading recovery.}
\label{fig:sparsity_recovery}
\end{subfigure}

\end{figure*}
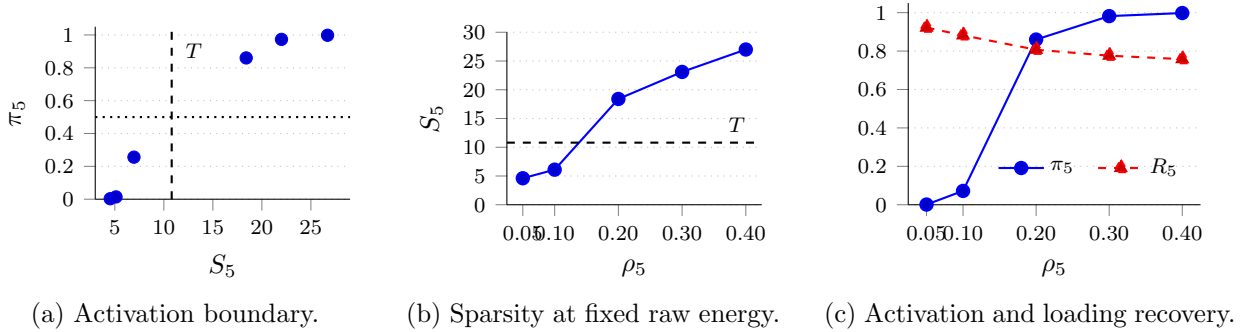

The transition in activation is accompanied by a pronounced increase
in standardised loading energy. Median $S_5$ increases from $4.53$ at
$c_5=0.20$ to $6.95$ at $c_5=0.30$, remaining below the
activation threshold of approximately $10.8$. At $c_5=0.35$,
however, $S_5$ increases to $18.4$ and clearly exceeds the threshold,
while $T$ itself remains approximately constant. Thus, the empirical
transition from shrinkage to activation is consistent with the factor
level evidence mechanism described in
Section~\ref{sec:ActivationBoundary}. This transition is illustrated
in Figure~\ref{fig:weak_factor_recovery}(a), where activation rises
sharply once the standardised loading energy moves beyond the
activation boundary. The null configuration is also well identified,
with $\Pr(H^+=4\mid\bm Y)=0.988$ at $c_5=0$.

\paragraph{Sparsity.}
Keeping the true raw loading energy fixed does not make activation
invariant to sparsity. When factor $5$ is supported on only $5\%$ of
the variables, median activation is essentially zero,
$\pi_5=0.001$, increasing to $0.072$ at $\rho_5=0.10$, $0.860$ at
the baseline $\rho_5=0.20$, and values close to one for
$\rho_5\geq0.30$. The corresponding standardised loading energy
increases from $S_5=4.61$ at $\rho_5=0.05$ to $27.0$ at
$\rho_5=0.40$, despite the raw loading energy
$\|\bm\lambda_{05}\|^2$ being held fixed. Figure~
\ref{fig:weak_factor_recovery}(b) shows that the change in support
size therefore moves the factor from below to well above the
activation boundary through its effect on the learned standardised
loading energy.

At the same time, loading recovery exhibits the opposite pattern.
It is highest for the sparsest configuration, with $R_5=0.922$ at
$\rho_5=0.05$, and decreases monotonically to $0.758$ at
$\rho_5=0.40$. As shown in Figure~
\ref{fig:weak_factor_recovery}(c), factor activation and loading
recovery therefore move in opposite directions as the support becomes
less sparse. In particular, the relative loading pattern across
variables can be recovered accurately even when the corresponding
factor is strongly favoured to belong to the spike. These results
show that factor activation and loading recovery capture distinct
aspects of weak factor recovery. Moreover, the large change in $S_5$
despite fixed raw loading energy shows that the effect of sparsity
cannot be explained by aggregate loading magnitude alone, but depends
on its interaction with the local and global shrinkage hierarchy.

\paragraph{Dimension.}
With the support fraction fixed at $0.20$, factor recovery improves
strongly with the number of observed variables. At $p=50$, median
activation is only $\pi_5=0.082$ and the posterior probability of the
correct factor dimension is $0.083$. At $p=100$, these increase to
$0.860$ and $0.849$, respectively, while for $p=200$ and $p=400$
the fifth factor is recovered with essentially unit posterior
probability.

Although the activation threshold increases with dimension, from
$T=5.53$ at $p=50$ to $T=42.0$ at $p=400$, the standardised loading
energy increases considerably faster, from $S_5=2.74$ to $104$.
Because the support proportion is fixed, the number of active
loadings grows proportionally with $p$, increasing the aggregate
signal available for factor activation. Hence, in this regime the
additional signal accumulated as $p$ increases more than compensates
for the dimensional contribution to the activation threshold, in
accordance with the dimensional regimes discussed in
Section~\ref{sec:Sparsity}.

\paragraph{Truncation.}
Increasing the truncation level raises the activation
threshold, from $T=10.6$ at $H=8$ to $T=11.6$ at $H=30$, as
predicted by the prior odds contribution $-\log(H/\alpha)$. The
effect on posterior activation, however, is not determined by this
threshold shift alone. Median standardised loading energy also
increases with $H$, from $S_5=13.8$ at $H=8$ to $30.7$ at $H=30$.
Correspondingly, median activation increases from $\pi_5=0.617$ at
$H=8$ to $0.860$ at $H=10$ and exceeds $0.97$ for $H\geq15$.

In contrast, loading recovery remains essentially unchanged across
the truncation levels, with $R_5$ close to $0.80$ throughout.
Recovery of the effective dimension also improves from
$\Pr(H^+=5\mid\bm Y)=0.601$ at $H=8$ to $0.849$ at $H=10$ and
approximately $0.96$--$0.97$ for $H\geq15$, with only a slight
decrease from $0.971$ at $H=20$ to $0.958$ at $H=30$. These results
illustrate the distinction between the direct contribution of
truncation to the ESP activation odds and its effect in the full
hierarchical model. Although increasing $H$ raises the conditional
activation threshold through $-\log(H/\alpha)$, the global and local
shrinkage parameters are themselves learned from the data, and the
resulting standardised loading energy $S_5$ changes with $H$. In the
present experiment, this posterior adaptation more than compensates
for the direct truncation penalty.

Across the experiments, factors $1$-$4$ are substantially more
stable than the deliberately weak fifth component. The greatest
posterior instability occurs close to the weak factor activation
boundary and, in the truncation experiment, at the smallest value
$H=8$. These are also the configurations in which the multi-chain
diagnostics indicate slower mixing of the factor selection
quantities. Detailed $\widehat R$, bulk-ESS, and tail-ESS diagnostics are reported in the Supplementary Material,
Section~\ref{app:MCMCDiagnostics}. The slower mixing of the factor-selection quantities noted at $H=8$  is plausibly related to this truncation level providing comparatively little overfitting margin relative to the true dimension $H^+=5$: with only three redundant columns available, there is less room for the sampler to reallocate factors between active and inactive states than at the larger truncation levels considered elsewhere, which can slow mixing of $H^+$, $\gamma_5$, and $S_5$.

\section{Conclusion}
\label{sec:concl}

We studied weak factor recovery in sparse Bayesian factor models under the ESP prior specification based on a variance mixture. We showed that posterior factor activation can be expressed as a competition between the evidence supplied by the loading column, summarised by the standardised loading energy $S_h$, and the ESP prior activation odds $\alpha/H$. This representation yields an explicit activation boundary and clarifies how signal strength, sparsity, the number of observed variables $p$, and the truncation level $H$ affect the distinction between active and redundant factors.

The simulation study supports this theoretical mechanism. Weak factors exhibit a clear transition from shrinkage to recovery as their standardised loading energy crosses the activation boundary. Factors with the same raw loading energy can nevertheless have substantially different activation probabilities when their signal is distributed over supports of different sizes, because sparsity affects the learned global and local shrinkage parameters and hence $S_h$. We also find that the relative loading pattern across variables can be recovered accurately even when the corresponding factor is strongly favoured to belong to the spike, showing that loading recovery and factor activation capture distinct aspects of recovery. Increasing $p$ can improve recovery when the support grows proportionally with dimension, whereas increasing $H$ raises the conditional activation threshold but does not necessarily reduce recovery, since posterior adaptation of the global and local shrinkage parameters can offset the direct truncation penalty. The results presented here pertain to the variance based ESP specification considered in this paper.

The activation mechanism studied here also highlights an important distinction between the exchangeable ESP prior and other shrinkage priors for factor models. Under the present ESP prior, the prior activation odds $\alpha/H$ are shared across all candidate factors, so that, conditional on the global and local shrinkage parameters, the activation boundary does not depend on the column index. This differs from the multiplicative gamma process prior of \citet{BhattacharyaDunson2011}, which induces progressively stronger continuous shrinkage through a cumulative product of gamma variables and has no discrete allocation indicator directly analogous to $\gamma_h$. The cumulative shrinkage process prior of \citet{Legramanti2020} is more closely related to the present specification through its spike-and-slab formulation, but its prior probability of shrinkage increases with the column index. An analogous activation boundary under CUSP would therefore generally involve index dependent prior odds rather than the common odds $\alpha/H$ arising under the exchangeable ESP. Deriving comparable weak factor recovery criteria under these alternative priors and evaluating them under the same weak factor scenarios considered here would provide a natural way to assess the role of exchangeability in weak factor recovery.

\bibliography{References}

\clearpage
\appendix
\numberwithin{equation}{section}

\begin{center}
{\Large\bfseries Supplementary Material for}\\[0.5em]
{\Large\bfseries \lq \lq Weak Factor Recovery in Sparse Bayesian Factor Models under Exchangeable Shrinkage Process Prior\rq \rq}
\end{center}

\section{Derivations for the posterior factor activation mechanism}
\label{app:PosteriorActiv}

This section derives the marginal distribution of a loading column underlying the conditional Bayes factor in  Section~\ref{sec:PosteriorActiv}. We use the following inverse gamma shape-scale parameterisation
$$
p(\theta)
=
\frac{b^a}{\Gamma(a)}
\theta^{-(a+1)}
\exp\left(-\frac{b}{\theta}\right),
\qquad
\theta>0.
$$

For factor $h$, let
$\bm\lambda_h=(\lambda_{1h},\ldots,\lambda_{ph})^\top$,
$\bm\phi_h=(\phi_{1h},\ldots,\phi_{ph})^\top$, and define
$\bm\Sigma_h=\kappa\,\mathrm{diag}(\bm\phi_h)$.
Conditional on the factor-specific variance $\theta_h$,
$$
\bm\lambda_h
\mid
\theta_h,\kappa,\bm\phi_h
\sim
N_p(\bm 0,\theta_h\bm\Sigma_h),
$$
with density
$$
p(\bm\lambda_h
\mid
\theta_h,\kappa,\bm\phi_h)
=
(2\pi)^{-p/2}
|\bm\Sigma_h|^{-1/2}
\theta_h^{-p/2}
\exp\left(
-\frac{S_h}{2\theta_h}
\right),
$$
where
$$
S_h
=
\bm\lambda_h^\top
\bm\Sigma_h^{-1}
\bm\lambda_h
=
\sum_{i=1}^p
\frac{\lambda_{ih}^2}{\kappa\phi_{ih}}.
$$

Under component $g\in\{0,1\}$ of the ESP prior,
$\theta_h\mid\gamma_h=g\sim IG(a_g,b_g)$. Integrating out
$\theta_h$ gives
\begin{align}
m_g(\bm\lambda_h)
&=
p(\bm\lambda_h
\mid
\gamma_h=g,\kappa,\bm\phi_h)
=
\frac{
(2\pi)^{-p/2}
|\bm\Sigma_h|^{-1/2}
b_g^{a_g}
}{
\Gamma(a_g)
}
\int_0^\infty
\theta_h^{-(a_g+p/2+1)}
\exp\left\{
-\frac{b_g+S_h/2}{\theta_h}
\right\}
d\theta_h
\nonumber\\
&=
(2\pi)^{-p/2}
|\bm\Sigma_h|^{-1/2}
\frac{
b_g^{a_g}
\Gamma(a_g+p/2)
}{
\Gamma(a_g)
\left(b_g+S_h/2\right)^{a_g+p/2}
}.
\label{eq:app_marginal_density}
\end{align}

This is the density of a multivariate Student-$t$ distribution with
$2a_g$ degrees of freedom, location $\bm 0$, and scale matrix
$(b_g/a_g)\bm\Sigma_h$. Hence,
$$
\bm\lambda_h
\mid
\gamma_h=g,\kappa,\bm\phi_h
\sim
t_{2a_g}\left(
\bm 0,
\frac{b_g}{a_g}
\kappa\,\mathrm{diag}(\bm\phi_h)
\right).
$$

\section{Properties of the factor activation boundary}
\label{app:ActivationBoundary}

This section provides additional derivations concerning the activation
boundary and the effects of the truncation level $H$ and the number of
observed variables $p$ discussed in
Section~\ref{sec:ActivationBoundary}.

\subsection{Existence and uniqueness of the activation threshold}

Under the monotonicity condition discussed in
Section~\ref{sec:ActivationBoundary}, the activation threshold $T$
satisfies
$$
BF_h(T,p)=\frac{H}{\alpha}.
$$
To establish existence, write
$$
BF_h(S_h,p)
=
C(p)
\frac{
\left(b_0+S_h/2\right)^{a_0+p/2}
}{
\left(b_1+S_h/2\right)^{a_1+p/2}
},
$$
where
$$
C(p)
=
\frac{
b_1^{a_1}\Gamma(a_1+p/2)\Gamma(a_0)
}{
b_0^{a_0}\Gamma(a_0+p/2)\Gamma(a_1)
}.
$$
As $S_h\rightarrow\infty$,
\begin{align}
BF_h(S_h,p)
&\sim
C(p)
\left(\frac{S_h}{2}\right)^{a_0-a_1}.
\label{eq:app_BF_asymptotic}
\end{align}
Thus, when $a_0>a_1$,
$$
BF_h(S_h,p)\rightarrow\infty
\qquad
\text{as }
S_h\rightarrow\infty.
$$
If in addition
$$
BF_h(0,p)<\frac{H}{\alpha},
$$
strict monotonicity and continuity imply that there exists a unique
positive threshold $T$ satisfying
$$
BF_h(T,p)=\frac{H}{\alpha}.
$$
For $S_h>T$, the posterior odds favour the slab, whereas for $S_h<T$,
they favour the spike. If instead
$$
BF_h(0,p)\geq\frac{H}{\alpha},
$$
the slab is already favoured at $S_h=0$ and, under global monotonicity,
remains favoured for all $S_h\geq0$.

\subsection{Effect of the truncation level}

To examine the effect of the truncation level on the activation
threshold, consider
$$
\log BF_h(T,p)
=
\log\left(\frac{H}{\alpha}\right).
$$
Treating $H$ as continuous for interpretative purposes and
differentiating implicitly gives
$$
\frac{dT}{dH}
=
\frac{
1/H
}{
\left.
\frac{\partial}{\partial S_h}
\log BF_h(S_h,p)
\right|_{S_h=T}
}.
$$
Whenever the Bayes factor is increasing at the activation boundary,
this derivative is positive. Thus, a larger truncation level raises
the standardised loading energy required for activation.

\subsection{Effect of the number of observed variables}

To characterise how the activation threshold varies with $p$, implicitly
differentiating
$$
\log BF_h(T,p)
=
\log\left(\frac{H}{\alpha}\right)
$$
with respect to $p$ gives
$$
\frac{dT}{dp}
=
-
\frac{
\left.
\frac{\partial}{\partial p}
\log BF_h(S_h,p)
\right|_{S_h=T}
}{
\left.
\frac{\partial}{\partial S_h}
\log BF_h(S_h,p)
\right|_{S_h=T}
}.
$$
Hence, even when the Bayes factor is increasing in $S_h$, the
direction in which the activation boundary moves with $p$ depends on
the sign of the direct dimensional effect.

\section{Simulation hyperparameters}
\label{app:SimulationHyperparameters}

The same prior hyperparameters are used throughout all simulation
experiments. We use the inverse gamma shape-scale parameterisation and the gamma shape-rate parameterisation. The values are summarised in Table~\ref{tab:simulation_hyperparameters}.

\begin{table}[h]
\centering
\small
\caption{Prior hyperparameters used in the simulation study.}
\label{tab:simulation_hyperparameters}
\begin{tabular}{lll}
\toprule
Parameter & Prior & Hyperparameters \\
\midrule
Residual variance $\xi_i^2$
& $IG(a_\xi,b_\xi)$
& $(3,2)$ \\

Global scale $\kappa$
& $IG(a_\kappa,b_\kappa)$
& $(3,1)$ \\

Local scale $\phi_{ih}\mid\beta_{\phi i}$
& $IG(a_\phi,\beta_{\phi i})$
& $a_\phi=1.5$ \\

Local rate $\beta_{\phi i}$
& $\mathrm{Gamma}(c_\phi,d_\phi)$
& $(1.5,1)$ \\

Spike $\theta_h\mid\gamma_h=0$
& $IG(a_0,b_0)$
& $(20,0.95)$ \\

Slab $\theta_h\mid\gamma_h=1$
& $IG(a_1,b_1)$
& $(3,1)$ \\

ESP concentration $\alpha$
& $\mathrm{Gamma}(a_\alpha,b_\alpha)$
& $(8,2)$ \\
\bottomrule
\end{tabular}
\end{table}

Under this specification, the prior means of the spike-and-slab
components are $0.05$ and $0.50$, respectively. 

\section{MCMC convergence diagnostics}
\label{app:MCMCDiagnostics}

For each simulated dataset and model configuration, posterior inference was based on four independently seeded MCMC chains. Convergence and sampling efficiency were assessed using the rank-normalised split-$\widehat R$, bulk effective sample size (bulk-ESS), and tail effective sample size (tail-ESS) proposed by
\citet{vehtari2021rank}. All diagnostics were computed using the
\texttt{rhat()}, \texttt{ess\_bulk()}, and \texttt{ess\_tail()}
functions in the R package \texttt{posterior}
\citep{buerkner2020posterior}.

\paragraph{Rank-normalised split-$\widehat R$.}
The $\widehat R$ diagnostic compares variation between MCMC chains with variation within chains, with values substantially larger than one indicating that the chains have not yet explored the same posterior distribution. We use the rank-normalised split-$\widehat R$ of \citet{vehtari2021rank}. Each chain is first split into two halves, and the pooled draws are transformed to normal scores using their ranks. A split-$\widehat R$ statistic is then computed from the transformed draws. To increase sensitivity to differences in scale between chains, the same calculation is also applied to folded draws, obtained from absolute deviations from the pooled median. The reported diagnostic is the larger of the rank-normalised split-$\widehat R$ and the rank-normalised folded split-$\widehat R$.

For $M$ split chains of length $L$, let $W$ denote the average within chain variance and $B$ the between chain variance. The corresponding variance estimate is
$$
\widehat{\mathrm{var}}^+
=
\frac{L-1}{L}W+\frac{1}{L}B,
$$
and the split diagnostic is
$$
\widehat R
=
\sqrt{
\frac{\widehat{\mathrm{var}}^+}{W}
}.
$$
For the rank-normalised diagnostic, this calculation is performed on the rank-transformed draws and on their folded counterparts, with the larger of the two values reported.

\paragraph{Bulk effective sample size.}
The effective sample size measures the number of independent posterior draws that would contain approximately the same amount of information as the autocorrelated MCMC sample. Bulk-ESS is computed from the rank-normalised split chains and measures sampling efficiency in the central part of the posterior distribution. A small bulk-ESS therefore indicates substantial autocorrelation or slow mixing even when the nominal number of retained draws is large.

\paragraph{Tail effective sample size.}
Tail-ESS assesses sampling efficiency in the tails of the posterior distribution. Following \citet{vehtari2021rank}, it is defined as the minimum of the effective sample sizes associated with the $5\%$ and $95\%$ posterior quantiles. It can therefore identify situations in which the chains mix adequately in the bulk of the distribution but explore its tails inefficiently.

For each fitted model, the diagnostics were calculated separately for the global shrinkage parameter $\kappa$, the ESP concentration parameter $\alpha$, and the effective factor dimension
$H^+=\sum_{h=1}^H\gamma_h$. When the fifth factor was present, posterior loading columns were first matched to the true factors at every retained draw using the same one-to-one assignment described in Section~\ref{sec:SimulationDesign}. For the posterior column $h_5$ matched to the true fifth factor, the activation indicator $\gamma_5$ and standardised loading energy
$$
S_5
=
\sum_{i=1}^p
\frac{\lambda_{ih_5}^2}{\kappa\phi_{ih_5}}
$$
were then monitored across chains.

Following \citet{vehtari2021rank}, we used
$\widehat R<1.01$ and $\mathrm{ESS}_{\mathrm{bulk}}\geq400$
as diagnostic benchmarks, with the latter corresponding to approximately $100$ effective draws per chain for the four-chain analysis. For the continuous monitored quantities
$\kappa$, $\alpha$, and $S_5$, we additionally considered
$\mathrm{ESS}_{\mathrm{tail}}\geq400$.
Tail-ESS was not used as a diagnostic benchmark for the discrete quantities $H^+$ and $\gamma_5$.

\paragraph{Diagnostic results.}
The ESP concentration parameter $\alpha$ showed consistently stable sampling across all simulation configurations, with all monitored fits satisfying $\widehat R<1.01$ and both bulk-ESS and tail-ESS exceeding $400$. Mixing was slower for the global shrinkage parameter $\kappa$, which was the most frequent source of diagnostic values outside these benchmarks. This was particularly apparent in some higher-dimensional and larger truncation configurations, even when the factor selection quantities themselves were stable.

The factor selection quantities $H^+$, $\gamma_5$, and $S_5$ showed their slowest mixing primarily in configurations close to the weak factor activation boundary. In the signal strength experiment, this behaviour was most pronounced at $c_5=0.30$ and remained visible at the baseline value $c_5=0.35$, whereas substantially weaker or stronger factors showed more stable sampling. A similar pattern appears in the truncation experiment. At $H=8$, where $S_5$ lies relatively close to the activation threshold and posterior activation remains uncertain, these quantities also show slower mixing. As $H$ increases and $S_5$ moves further above the threshold, their diagnostics become substantially more stable. This pattern is consistent with greater posterior uncertainty near the spike-slab activation boundary, where posterior mass may be distributed across alternative factor activation states. These diagnostic benchmarks were used to characterise sampling behaviour rather than to exclude individual simulation replicates.

\end{document}